\documentstyle[epsfig]{elsart}
\begin{document}
\hyphenation{ra-dio sig-nal a-zi-muth com-pu-ters 
sen-si-ti-vi-ty de-ve-lopment 
e-ner-gy ma-xi-mum vo-lumes}
\begin{frontmatter}
\title{
Progress in Air Shower Radio Measurements: 
Detection of Distant Events 
}
LOPES Collaboration
\author[aaa]{W.D.~Apel},
\author[iii]{T.~Asch},
\author[aaa]{A.F.~Badea}$^,$\footnote{corresponding author: 
Florin.Badea@ik.fzk.de}$^,$\footnote{on leave of absence from $^d$},
\author[bbb]{L.~B\"ahren},
\author[aaa]{K.~Bekk},
\author[ccc]{A.~Bercuci},
\author[ddd]{M.~Bertaina},
\author[eee]{P.L.~Biermann},
\author[aaa,fff]{J.~Bl\"umer},
\author[aaa]{H.~Bozdog},
\author[ccc]{I.M.~Brancus},
\author[ggg]{S.~Buitink},
\author[hhh]{M.~Br\"uggemann},
\author[hhh]{P.~Buchholz},
\author[bbb]{H.~Butcher},
\author[ddd]{A.~Chiavassa},
\author[fff]{F.~Cossavella},
\author[aaa]{K.~Daumiller},
\author[ddd]{F.~Di~Pierro},
\author[aaa]{P.~Doll},
\author[aaa]{R.~Engel},
\author[bbb,eee,ggg]{H.~Falcke},
\author[iii]{H.~Gemmeke},
\author[jjj]{P.L.~Ghia}, 
\author[kkk]{R.~Glasstetter},
\author[hhh]{C.~Grupen},
\author[aaa]{A.~Haungs},
\author[aaa]{D.~Heck},
\author[fff]{J.R.~H\"orandel},
\author[ggg]{A.~Horneffer},
\author[aaa]{T.~Huege},
\author[kkk]{K.H.~Kampert},
\author[hhh]{Y.~Kolotaev},
\author[iii]{O.~Kr\"omer},
\author[ggg]{J.~Kuijpers},
\author[ggg]{S.~Lafebre},
\author[aaa]{H.J.~Mathes},
\author[aaa]{H.J.~Mayer},
\author[aaa]{C.~Meurer},
\author[aaa]{J.~Milke},
\author[ccc]{B.~Mitrica},
\author[jjj]{C.~Morello},
\author[ddd]{G.~Navarra},
\author[aaa]{S.~Nehls},
\author[ggg]{A.~Nigl},
\author[aaa]{R.~Obenland},
\author[aaa]{J.~Oehlschl\"ager},
\author[aaa]{S.~Ostapchenko}$^,$\footnote{on leave of absence from Moscow State 
University, 119899 Moscow, Russia},
\author[hhh]{S.~Over},
\author[ccc]{M.~Petcu},
\author[ggg]{J.~Petrovic},
\author[aaa]{T.~Pierog},
\author[aaa]{S.~Plewnia},
\author[aaa]{H.~Rebel},
\author[mmm]{A.~Risse},
\author[aaa]{M.~Roth},
\author[aaa]{H.~Schieler},
\author[ccc]{O.~Sima},
\author[ggg]{K.~Singh},
\author[fff]{M.~St\"umpert},
\author[ccc]{G.~Toma},
\author[jjj]{G.C.~Trinchero},
\author[aaa]{H.~Ulrich},
\author[aaa]{J.~van~Buren},
\author[hhh]{W.~Walkowiak},
\author[aaa]{A.~Weindl},
\author[aaa]{J.~Wochele},
\author[mmm]{J.~Zabierowski},
\author[eee]{J.A.~Zensus},
\author[hhh]{D.~Zimmermann}
\address[aaa]{Institut f\"ur Kernphysik, Forschungszentrum Karlsruhe,
76021 Karlsruhe, Germany}
\address[iii]{Institut f\"ur Prozessdatenverarbeitung und Elektronik, 
Forschungszentrum Karlsruhe, 76021 Karlsruhe, Germany}
\address[bbb]{ASTRON, 7990 AA Dwingeloo, The Netherlands}
\address[ccc]{National Institute of Physics and Nuclear Engineering, 7690 
Bucharest, Romania}
\address[ddd]{Dipartimento di Fisica Generale dell' Universita, 10125 Torino, 
Italy}
\address[eee]{Max-Planck-Institut f\"ur Radioastronomie, 53121 Bonn, Germany}
\address[fff]{Institut f\"ur Experimentelle Kernphysik, Universit\"at Karlsruhe,
76021 Karlsruhe, Germany}
\newpage
\address[ggg]{Dpt. Astrophysics, Radboud University, 6525 ED Nijmegen,
The Netherlands}
\address[hhh]{Fachbereich Physik, Universit\"at Siegen, 57072 Siegen, Germany}
\address[jjj]{Istituto di Fisica dello Spazio Interplanetario, INAF, 10133 
Torino, Italy}
\address[kkk]{Fachbereich C $-$ Physik, Universit\"at Wuppertal, 42097 
Wuppertal, Germany}
\address[mmm]{Soltan Institute for Nuclear Studies, 90950 Lodz, Poland}

\begin{abstract} 
Data taken during half a year of operation of 10 LOPES antennas 
(LOPES-10), triggered by EAS observed with 
KASCADE-Grande have been analysed. 
We report about the analysis of correlations of radio signals 
measured by LOPES-10 with extensive air shower
events reconstructed by KASCADE-Grande, including   
shower cores at large distances.
The efficiency of detecting radio signals induced by air showers 
up to distances of $700\,$m 
from the shower axis has been investigated. 
The results are discussed with special emphasis on the effects of the
reconstruction accuracy for shower core and arrival direction on the
coherence of the measured radio signal. 
In addition, the correlations of the radio pulse amplitude with the 
primary cosmic ray energy and with the 
lateral distance from the shower core are studied.
\end{abstract}
\end{frontmatter}

\section{Introduction}
In 1962 Askaryan~\cite{Askar61} predicted  that 
extensive air showers (EAS) should generate coherent radio emission. 
A few years later, in 1965, the phenomenon has been experimentally 
discovered by observing a radio pulse generated during the 
EAS development at $44\,$MHz~\cite{Jelly65}. 
In a review by Allan~\cite{Allan71} these early studies
were summarized.
In the pioneering work of Askaryan coherent Cherenkov 
radiation of the charge-excess was considered as the process 
responsible for the EAS radio emission. 
However, the low matter density in the Earth's Atmosphere
and the existence of the Earth's magnetic field allow 
an alternative process for the origin of the radio signals.
In an approach based on coherent geosynchrotron 
radiation~\cite{Gorha03}, electron-positron pairs generated in the 
shower development gyrate in the Earth's magnetic field and emit 
radio pulses by synchrotron emission.
Detailed analytical studies~\cite{Huege03} and Monte-Carlo
simulations~\cite{2Hueg05} predict relevant radio 
emission at frequencies of $10$ to a few hundred MHz. 
During the shower development the electrons are concentrated in a  
shower disk, with a thickness of a few meters.
This leads to a coherent emission at low frequencies up 
to $100\,$MHz, where the wavelength is larger than this thickness.
For showers above a certain threshold energy 
one expects a short, but 
coherent radio pulse of $10\,$ns to $100\,$ns duration
with an electric field strength increasing approximately linearly
with the primary energy of the cosmic particle 
inducing the air shower. I.e., one expects a quadratic increase 
of the received energy of the radio pulse with the primary particle 
energy. 
In addition, the geosynchrotron emission process is expected to be  
dominant for radio emission during the cosmic ray air shower 
development.
Recently, measurements using 
the LOPES experiment~\cite{Falck05} support these predictions. 

A series of recent papers 
(e.g.~\cite{Falck05,Horne04,Ravel04}) 
report about promising experimental results in detecting radio 
emission in coincidence with air shower events measured by 
particle detectors. 
A rather unique opportunity for 
calibrating and understanding the radio emission in EAS is provided 
by LOPES, which is located at the site of the KASCADE-Grande  
experiment~\cite{Navar04,Anton03}. 
The KASCADE-Grande experiment installed at the Forschungszentrum 
Karlsruhe is a multi-detector setup which allows precise
estimations of charged particle EAS observables in the primary 
energy range of $10^{14}-10^{18}\,$eV. 
LOPES-10 is an array of 10 dipole antennas 
placed inside the particle detector array of KASCADE-Grande.
Recently, LOPES-10 was extended with 20 additional antennas to
LOPES-30~\cite{Nehl05}. 

The LOPES-10 data set is subject of various analyses addressing 
different scientific questions. 
With a sample asking for high quality events 
the proof of principle for detection of air showers in the radio 
frequency range was made~\cite{Falck05}. 
With events falling inside the original smaller KASCADE array 
basic correlations of the radio signals with 
shower parameters were shown~\cite{Horne05a}.
Further interesting features are currently being investigated 
with a sample of very inclined showers~\cite{Petro05} and with 
a sample of events measured during thunderstorms~\cite{Buiti05}.  
The focus of the present paper is to 
display some general features deduced 
from first measurements of the LOPES experiment 
for events with primary energies 
above $5\cdot10^{16}\,$eV which are observed by 
the large KASCADE-Grande array and LOPES in coincidence,
including shower cores at large distances.
Special emphasis is put on effects of the applied radio 
reconstruction procedures, on detection efficiency, and on 
features of the measured radio signal dependencies on parameters of 
the primary cosmic particle such as the primary energy 
and distance of the antennas from the shower axis.

\section{Data processing}

\subsection{Experimental situation}

The KASCADE-Grande experiment~\cite{Navar04,Anton03} 
(Fig.~\ref{experiment}, left panel) 
observes EAS in the primary energy range from $100\,$TeV 
to $1\,$EeV. 
It enables multi-parameter measurements of a large number of 
observables of the three main EAS components: 
electron$/\gamma$ component, muons, and hadrons.
The main detector parts are the original KASCADE array, and 
the Grande detector stations (Fig.~\ref{experiment}, left panel).  
The KASCADE array consists of 252 scintillator
detector stations and measures the 
electromagnetic and muonic components with $5\,$MeV and $230\,$MeV 
energy thresholds, respectively. 
The array is organized in 16 quadratic clusters, where the outer
12 clusters contain electron (unshielded) and muon (shielded)
detectors, while the inner four clusters contain 
electron$/\gamma$- detectors, only.
The 37 stations of the Grande array,   
covering an area of approx. $0.5\,$km$^2$, take data in 
coincidence with KASCADE and allow to reconstruct showers with 
distances between 
shower core and the LOPES-10 antennas up to $850\,$m.
The Grande array is triggered by a coincidence of seven neighboring 
stations. 
The present analysis uses data of the Grande array for 
reconstructing basic shower parameters: location and direction of
the shower axis and the shower size. In addition, data of the 
original KASCADE array is used to reconstruct the muon number 
of the individual showers. 

LOPES is an array of radio antennas, which  
has been installed on the site of the KASCADE-Grande 
experiment in order to demonstrate the feasibility of EAS 
radio measurements.
LOPES is based on prototype developments for the 
Low-Frequency-Array (LOFAR)~\cite{LOFAR04}.
In the current status LOPES 
({\bf LO}far {\bf P}rototyp{\bf E} {\bf S}tation) 
operates 30 short dipole radio 
antennas (LOPES-30); the present analysis uses only 
data of the first 10 antennas forming LOPES-10 
(Fig.~\ref{experiment}, left part). 
The antennas, positioned in the original KASCADE array, 
operate in the frequency range of 
$40-80\,$MHz and are aligned in east-west direction, i.e.
they are sensitive to the linear 
east-west polarized component of the radiation. 
The read out window for each antenna is $0.8\,$ms wide, 
centered around the trigger received from the KASCADE array. 
The sampling rate is $80\,$MHz.
The logical condition for the LOPES-trigger 
is at least 10 out of the 16 KASCADE clusters to be fired. 
This corresponds to primary energies above $\approx 10^{16}\,$eV; 
such showers are detected at a rate of $\approx 2$ per minute. 
The LOPES-10 experiment and its electronics 
is described in more detail in~\cite{Horne04,Horne05a}.

\subsection{Selection of candidate events}

\begin{figure}[ht]
\begin{center}
\includegraphics[width=13.cm]{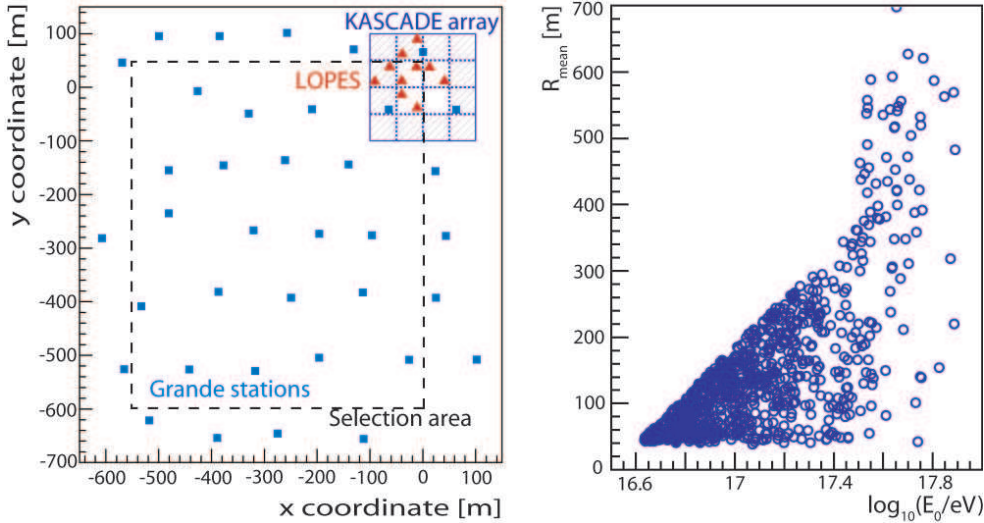}
\end{center}
\caption{Left panel: Sketch of the KASCADE-Grande experiment. 
The main detector components are the KASCADE array and the Grande
stations. The location of the 10 LOPES
radio antennas is also 
displayed. 
Right panel: Distribution of the selected candidate events 
(862 events) with respect to 
primary energy (obtained by Grande reconstruction) 
and mean distance of the shower axis to the radio antennas.
The selection is based on cuts of eqn.\ref{re0cut}, see text.}
\label{experiment}
\vspace*{0.3cm}
\end{figure}

A sample of 862 candidate events is selected
out of five months of LOPES-10 data 
taken in coincidence with the Grande array 
(Fig.~\ref{experiment}, right panel).
Selection criteria are \\
i) coincident measurements of the event by LOPES-10, 
KASCADE field array which has triggered LOPES, and Grande array; \\
ii) a successful reconstruction of the shower observables 
by the Grande array; \\
iii) zenith angle of the shower less than $50^\circ$; \\
iv) a geometrical cut to ensure the core position to be inside the 
fiducial area ($0.358\,$km$^2$) of the Grande array, 
see Fig.~\ref{experiment}, left panel; \\
v) and, to reduce the amount of data, an energy and 
distance cut is additionally applied which is motivated by 
Allan's formula, as explained below.\\
The historical measurements of the 1960s were compiled by 
Allan~\cite{Allan71} with the result that 
the pulse amplitude per unit bandwidth ($\epsilon_\nu$) of 
the radio signal induced by an EAS is described by the formula:
\begin{equation}
\epsilon_\nu=20 \cdot \left(\frac{E_0}{10^{17}{\rm eV}}\right) 
\cdot \sin{\alpha} \cdot \cos{\theta} \cdot
\exp\left(-\frac{R}{R_0(\nu,\theta)}\right)\,\,
\left[{\rm \frac{\mu V}{m\cdot MHz}}\right]
\label{allan}
\end{equation}
with $E_0$ the primary particle energy in eV, 
$\alpha$ the angle between shower axis and the geomagnetic field, 
$\theta$ the shower zenith angle, 
$R$ the antenna distance to the shower axis, 
and the scaling radius $R_0(\nu,\theta)$, which is in the range
of $30-300\,$m with $R_0=110\,$m at 
$55\,$MHz and $\theta<35^\circ$.\\ 
Compared to $R$ in the formula, 
for the present analysis $R_{\rm mean}$ is 
defined as the mean of the distances between the 10 LOPES 
antennas and the shower core position (reconstructed by Grande) 
in shower coordinates, 
i.e. in planes perpendicular to the axis direction.
The cut has been applied as
\begin{equation}
\log_{10}\left(\frac{E_0}{\rm eV}\right) > 
\log_{10}\left(\frac{E_{00}}{\rm eV}\right) + 0.4343\cdot 
\frac{R_{\rm mean}}{R_0}\,\,\,\,{\rm or}\,\,\,
\,\log_{10}\left(\frac{E_0}{\rm eV}\right)>17.5
\label{re0cut}
\end{equation}
with $E_{00}=10^{16.5}\,$eV and $R_0=160\,$m, i.e. weaker than 
Allan's scaling with radius. 
The threshold primary energy $E_{00}$ has been chosen based on the 
results from ref.~\cite{Falck05} and data reduction considerations. 
In this selection there are no conditions on the weather situation 
at the KASCADE-Grande site, 
i.e. environmental corrections were not applied.
There is a known effect of an amplification of the radio signal 
during thunderstorms~\cite{Buiti05}, but during the discussed 
measuring time this affects less than 3\% of the selected events. 

\subsection{Analysis procedures}

The Grande array measures the densities and arrival times of the 
charged particles, from which shower core position and arrival 
direction are reconstructed.
The reconstruction of the shower size $N_e$ is also based on data
of the Grande array, where the lateral distribution of the 
measured densities is described by a slightly modified 
NKG-function~\cite{Glass05}. 
The total muon number is obtained by a likelihood fit of the 
muon densities measured by the KASCADE muon detectors, 
located in the outer 192 stations of the KASCADE 
array~\cite{Buren05}.  
In order to select and investigate the candidate events, 
the primary energy has been roughly estimated from measured
and angular corrected 
electron and muon numbers by a linear combination where the 
parameters have been deduced from simulations 
with fixed energies and five different 
primary masses by means of a linear regression 
analysis~\cite{Glass05}. 
These preliminary Grande reconstruction procedures 
applied to the first year of measurements with Grande lead to 
accuracies of the shower core position and direction in 
the order of $10\,$m and $0.5^{\circ}$ with 
$68$\% confidence level for simulated proton and iron showers with
$>50\,$PeV primary energy and $22^{\circ}$ zenith 
angle. The energy resolution is estimated to be 
$\Delta E/E \approx 30$\% in the relevant energy range which is 
sufficient for the following considerations.

The main steps to process the measured LOPES radio raw signals 
of an individual air shower are the following 
(for a more detailed description see refs.~\cite{Horne05a,Horne06}):
\begin{enumerate}
\item Correction of instrumental delays by monitoring 
the relative phases of a TV transmitter in the 
measured frequency band. 
\item Correction of the data for a frequency dependent gain factor 
of all electronic components in the signal chain. 
This factor has been obtained by measuring appropriate 
amplifications and attenuations in a laboratory environment.
\item Removal of narrow band radio frequency interference (RFI). 
It occupies only a few channels in frequency space, while
a short time pulse (e.g. from an EAS) is spread over 
all frequency channels. 
So, by flagging the channels with RFI the background is 
significantly reduced without affecting the air shower pulse 
significantly.
\item The digital beam-forming. It consists of two steps: 
First, a time shift of the data according to the given direction 
is done and then the combination of the data is performed calculating
the resulting beam from all antennas. 
The geometrical delay (in addition to the instrumental delay 
corrections) by which the data is shifted, is the time difference 
of the pulse coming from the given direction to reach 
the position of the corresponding antenna compared to the reference 
position. 
This shift is done by multiplying a phase gradient in the frequency 
domain before transforming the data back to the time domain.
This step includes also a correction for the azimuth and zenith
dependence of the antenna gain. \\
To form the beam from the time shifted data, the data from each 
pair of antennas is multiplied time-bin by time-bin, the resulting 
values are averaged, and then the square root is taken while 
preserving the sign.
We call this the cross-correlation beam or CC-beam: 
\begin{equation}
CC(t) = {\rm sign}(S(t)) \sqrt{\frac{1}{N_p} \left| S(t) \right|} 
\hspace*{0.5cm}  {\rm with} \hspace*{0.5cm} 
S(t) = \sum_{i \not= j}^{N} s_i(t) \cdot s_j(t)  \hspace*{0.3cm} ,
\label{ccbeam}
\end{equation}
with $N$ the number of antennas, 
$N_p$ the number of antenna pairs, 
$s_i(t)$ the field strength of antenna $i$, 
and $t$ the time-bin index.\\
The radio wavefront of an air shower is not expected to arrive
as a plane wave on the ground, it should have some curvature. 
During the reconstruction procedures the radius of this curvature 
is taken into account by iterating a free
parameter until the CC-beam is maximal. 
To some extent, however, the obtained value of this 
free parameter is degenerated by the uncertainties in the shower
direction and due to the fact that the signal is generated in
an extended and not point-like source.
\item Quantification of the radio parameters:
Due to the filtering of low and
high frequencies, the response of the analog electronics to a 
short pulse is an oscillation over a short time. 
Sampling such a signal with an ADC gives
a certain fine structure inside the pulse that is not part of 
the original pulse but is caused mainly by the filter. 
To suppress this fine structure 
the data is smoothed by block averaging over three samples in the 
time domain, where three was found as optimum value
not to broaden the pulse too much.
Although the shape of the resulting pulse (CC-beam) is not really 
Gaussian, fitting a Gaussian to the smoothed data gives
a robust value for the peak strength, which is defined as the
height of this Gaussian. 
The error of the fit results gives also a first estimate of the 
uncertainty of this parameter. 
The uncertainties dependent 
on the accuracy of the core and direction estimate used as 
input to the beam-forming are not considered in the following
analyses.
The obtained value for $\epsilon_\nu$, which is the 
measured amplitude divided by the effective bandwidth,
will be given in units of $[\mu$V/m$\cdot$MHz$]$, but has to be
multiplied by an unknown quantification factor $A$ due to the
lack of an absolute calibration~\cite{Horne06}.
\item Identification of good events:
Not every selected air shower is accompanied by
a radio pulse which is detectable by LOPES. There can also be
an incoherent noise peak which is as high as a peak induced by the
air shower, even in the formed beam. 
One can therefore not select events with 
air shower pulses just by the 
height of the fitted Gaussian, but has to classify events in a 
further step. 
The criteria for this selection are:
existence of a pulse, coherence of the pulse, 
expected position of the pulse in time,
and an approximately uniform pulse height in all antennas. 
Up to now these criteria are verified by hand, where 
the uncertainties are larger at low signal to noise ratios, i.e. 
at the threshold of detection.
\end{enumerate}

\subsection{Search for maximum radio coherence}

The observed radio signal is expected to be sensitive 
to characteristics of the primary particle inducing the
air shower like primary energy and mass.
A crucial element of the detection method is the digital 
beam-forming which allows to place a beam in the 
direction of the cosmic ray event. 
Therefore, the measured signal is also expected to be sensitive to
the core position, to the shower direction, and to the 
curvature of the emitted wave front.
To investigate such intrinsic capabilities 
of the beam-forming in case of LOPES-10 a simple simulation has 
been performed based on analytical calculations.

\begin{figure}[ht]
\includegraphics[width=14.0cm]{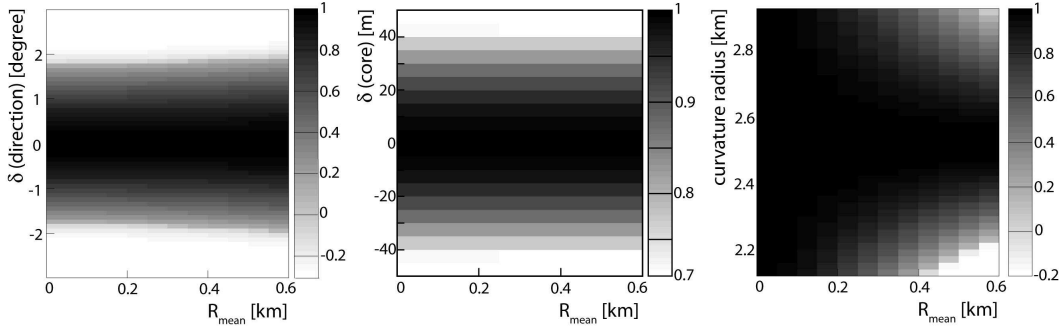}
\caption{Analytically estimated sensitivity of the 
(cross correlation) CC-beam to the variation of the 
shower direction $\delta$(direction), the shower core 
position $\delta$(core), and the radius of curvature 
of the radio front in dependence of $R_{\rm mean}$
for the LOPES-10 configuration, assuming a monochromatic 
point source emitting at $60\,$MHz.}
\label{sensitivity}
\vspace*{0.3cm}
\end{figure}

For a monochromatic point source radiating at 
$60\,$MHz, placed at a position of $2.5\,$km 
along the vertical shower axis from the ground, the 
sensitivity of the CC-beam estimator (eqn.~\ref{ccbeam})
has been calculated.
Fig.~\ref{sensitivity} gives an impression of the relation of 
the CC-beam to the shower direction, the shower core position 
and the radius of curvature of the radio front for the 
geometrical layout of LOPES-10. 
The $\delta$-values denote the differences between the true 
position/direction of the radio source and the assumed input 
values for calculating the CC-beam.
The z-axes in the figure are normalized to the 
value of the CC-beam estimator in case of `perfect' coherence, 
i.e. no uncertainties in the position of the radio 
point source ($\delta$(direction)=0, 
$\delta$(core)=0, and radius of curvature$=2.5\,$km). 
Only one parameter is varied at a time, where the other two 
are set to the true value.
The calculations show that one looses 
(z-value is decreased by 20\%) coherence if the start 
parameters for the CC-beam estimation in the direction are 
reconstructed with an error of more than $0.8^\circ$, 
and the core position with an error of more  
than $25\,$m, respectively. 
In contrast to direction and core, the distance   
of the shower axis from the antennas is relevant for 
the radius of curvature.
The further away from the axis, 
the more important the precision of 
the radius parameter becomes. \\
On the other hand the results shown in Fig.~\ref{sensitivity} can 
be interpreted as the intrinsic resolution of the antenna system, 
i.e. LOPES-10 has a limit in direction resolution 
of $0.8^\circ$ and in core resolution of $25\,$m, and the further 
away the shower axis is the better the resolution gets 
as the uncertainty in the radius estimate decreases. \\
The calculations have also shown that going to lower frequencies 
the better coherence improves the primary energy
estimation, where measuring with more antennas (LOPES-30) 
improves the reconstruction accuracy of the shower geometry. 

As shown in Fig.~\ref{sensitivity}, the procedure of time shifting 
the radio signals is relatively safe when the 
shower parameters for core and axis  
are reconstructed with high accuracy, 
i.e. provided by the reconstruction of data 
taken with the original KASCADE field array. 
Due to the high sampling area the 
accuracy of the core position 
and direction is good enough to 
obtain satisfying coherence of the radio signals. 
Of course, this is valid only for showers with cores inside 
KASCADE. 
A shower reconstruction using data from the Grande 
array is required for shower cores outside KASCADE. 
The Grande stations cannot assure an accuracy comparable 
with the original KASCADE array. 
This leads to events whose reconstructed radio signals  
do not fulfill the requirements to qualify as detected 
in the radio channel.
Therefore, a so-called optimized beam-forming is performed, 
which searches for maximum coherence by varying the core and 
the direction around the values provided by the 
Grande reconstruction. 
I.e., the beam-forming procedures 
- steps 4-6 as described above -
are repeated 50 times per 
shower, where the core and direction are randomly chosen
inside the parameter space given by the KASCADE-Grande 
reconstruction accuracy.
A more detailed discussion of the optimized beam-forming  
can be found in~\cite{Badea06}.

\begin{figure}[ht]
\vspace*{0.1cm}
\begin{center}
\includegraphics[width=12.cm]{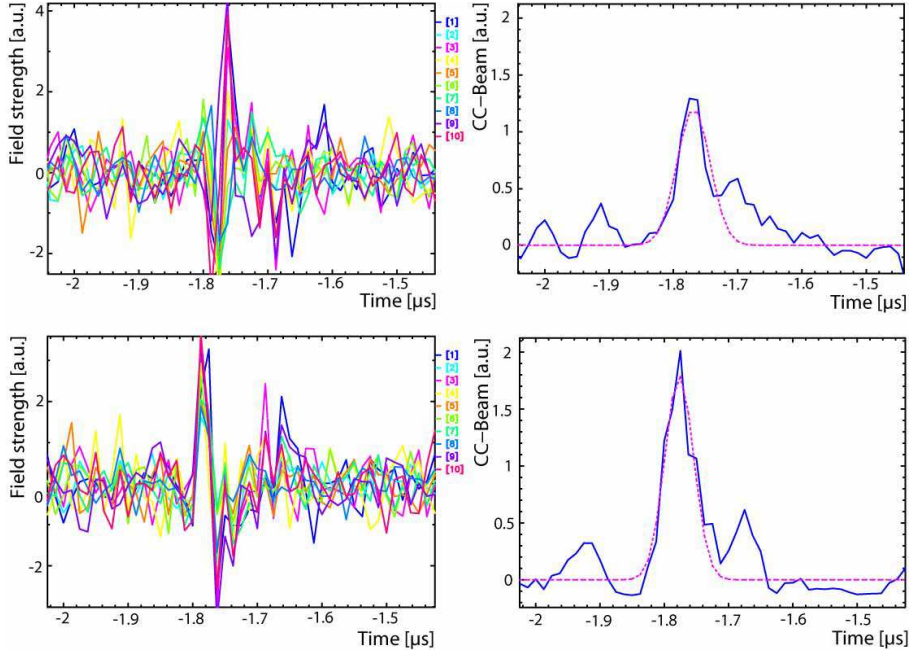}
\end{center}
\caption{Event example:
Upper panels: Signals of the individual antennas and result of the 
beam-forming (full line: CC-beam; dotted line: Gaussian fit) 
based on shower observables reconstructed by Grande; 
Lower panels: Signals of the individual antennas and result of the 
optimized beam-forming in order to maximize the radio coherence.}
\label{691}
\vspace*{0.3cm}
\end{figure}

Fig.~\ref{691} shows as an example the result of such an 
optimized beam-forming for an event with a medium distance 
between shower axis and radio antennas. 
In the upper part of the figure the raw 
time-series of the 10 antennas and the corresponding CC-beam 
including the fit are shown, obtained by using the 
Grande reconstructed parameter set. 
\begin{table}[ht]
\caption{Shower observables reconstructed by Grande alone and 
obtained after optimized
beam-forming of LOPES-10 data for the EAS radio event displayed in 
Fig.~\ref{691}.
The values of the shower core are given in 
KASCADE-Grande coordinates (Fig.~\ref{experiment}), 
the mean distance is the mean of the 
distances from the shower axis to the individual antennas.}
\label{691and231}
\begin{footnotesize}
\begin{center}
\begin{tabular}{|c||c|c|c|c|c|c|} \hline 
Observable &{$\rm \phi$}&{$\rm \theta$}&$\rm X_{core}$&$\rm Y_{core}$
&$R_{\rm mean}$&$E_0$ \\
 \hline 
Grande&289.5$^\circ$&  41.1$^\circ$& -66.0 m&-124.0 m& & \\
LOPES& 292.2$^\circ$& 40.6$^\circ$& -64.4 m&-123.0 m &
\raisebox{1.5ex}[1.5ex]{149.8 m}& 
\raisebox{1.5ex}[1.5ex]{$4 \cdot 10^{17}\,$eV}\\
                             \hline			     
\end{tabular}
\vspace*{0.3cm}
\end{center}
\end{footnotesize}
\end{table} 
The lower part shows the same event by choosing those 
starting parameters for the beam-forming which led to the 
maximum coherence, i.e. the highest radio pulse.
An increase of 50\% is seen in the CC-beam estimator after the 
optimized beam-forming. 
Table~\ref{691and231} compares the shower parameter values 
and their changes after achieving maximum coherence. 

In the sample of the candidate events there are also events 
which have a very short mean distance to the antennas.
Due to this small distance, the particle densities detected 
by the KASCADE array are relatively high inducing a lot of 
radio noise (incoherent signals, but still visible in the 
CC-beam estimation). After optimized beam-forming
these incoherent signals cancel out at the CC-beam. 
Events hitting the KASCADE array can 
also be reconstructed with KASCADE data independent of the 
Grande reconstruction, but with better accuracy.
It was found~\cite{Badea06} that for these events 
the two independently 
estimated parameter sets (Grande plus LOPES after the optimized 
beam-forming and KASCADE reconstruction)
are in better agreement than KASCADE with Grande results alone.
Therefore, the optimized beam-forming gives the possibility 
to improve the accuracy of shower parameters by including 
the radio information in the reconstruction of the shower 
with Grande. 
This concerns the core position and shower direction directly, 
then iteratively with the new values the reconstruction 
of the shower sizes and therefore also the primary energy and 
mass estimation. 

For events with very large distances, i.e. around or above 
$400\,$m, we also found a significant improvement of the
CC-beam detectability by applying the optimized beam-forming. 
For the reconstruction of such events the expected time shifts,
calculated by use of the known core distance, have to be taken into 
account to find the correct coherent peak.
Such shifts serve as an additional constraint on the analysis 
procedure, especially for events with large distance 
to the antennas~\cite{Badea06}.

After applying the optimized beam-forming procedures the 
obtained shower parameters are compared to the original 
(Grande reconstructed) parameters.
For the angle a mean shift of $2.3^\circ$ appears, which is
still reasonable if we assume an uncertainty of roughly $1^\circ$
for the present state of Grande reconstruction~\cite{Glass05} and 
also $\approx 1^\circ$ for LOPES-10~\cite{Falck05}. 
The mean shift in core position results to $15\,$m 
which is also not far off from the reconstruction uncertainties,
but for the individual events a systematic shift of the 
cores in direction to the antenna center was recognized. 
However, with the available statistics and still partly 
preliminary calibrations for both the Grande and LOPES detectors, 
it is premature to investigate any possible systematic shift 
between the `radio' and the `particle' axis of the shower.

\section{Results}

\subsection{Efficiency of the radio detection}

As shown in the previous section, the search for a coherent  
radio signal is very sensitive to the accuracies of reconstructing 
core position and shower direction. 
\begin{figure}[ht]
\begin{center}
\includegraphics[width=11.cm]{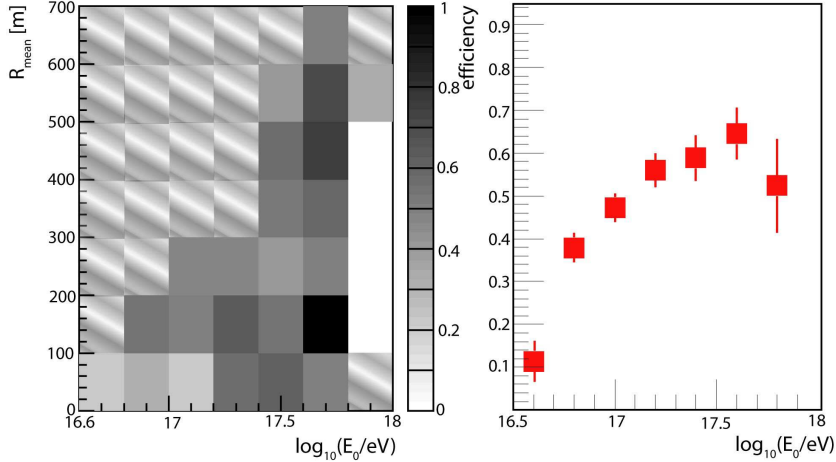}
\end{center}
\caption{Left panel: Efficiency of the radio detection after the 
optimized beam-forming (372 out of 862 candidate events). 
The hatched areas are bins where no candidates were selected
(Fig.~\ref{experiment}); 
Right panel: Efficiency for radio detection versus primary energy.}
\label{efficiency}
\vspace*{0.3cm}
\end{figure}
Due to the insufficient resolution of Grande, 
a lot of radio events fail in the first CC-beam reconstruction. 
Only 101 of the candidate events could be identified as radio showers.
After searching for maximum coherence by varying the core 
position and direction (i.e. optimized beam-forming), 
the number of detected radio events increases from 101 to
372 out of 862 candidate events 
(Fig.~\ref{efficiency}, left panel). 
The very small efficiencies visible in the lower-left part of 
this figure are caused by low primary energies as well 
as the high radio noise coming from the KASCADE particle 
detectors. The best efficiency is reached for showers with high 
energy but not too large distances, where the signal is 
still strong but the noise from the particle detectors weak.

One of the most interesting results of the current analysis is 
the presence of clear EAS radio events at more than
$500\,$m distance from the shower axis for primary energies 
below $10^{18}\,$eV.
Concerning the overall detection threshold 
an increasing efficiency with increasing primary 
energy reaching approximately 60\% for primary energies 
above $2\cdot10^{17}\,$eV is obtained with LOPES-10 
(Fig.~\ref{efficiency}, right panel).
\begin{figure}[ht]
\begin{center}
\includegraphics[width=7.cm]{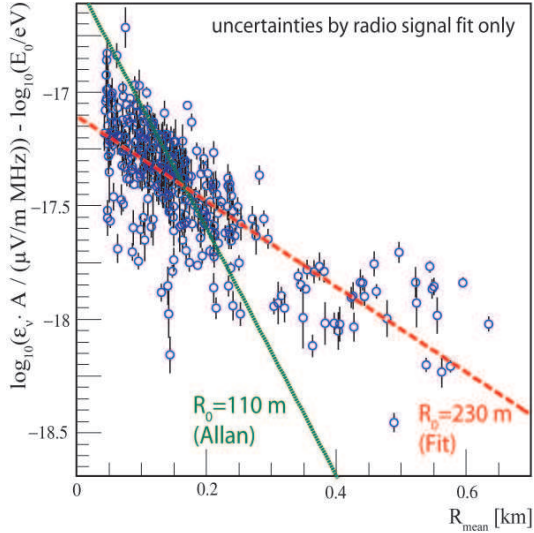}
\end{center}
\caption{Correlation of the pulse height corrected for primary energy 
with the mean distance of the shower axis to the radio antenna 
system. The lines show results of fits with an
exponential function with two free parameters (fit) and with fixed
scaling radius as suggested by Allan~\cite{Allan71}, respectively.}
\label{expomeanr}
\vspace*{0.3cm}
\end{figure}
An aggravating circumstance for missing detection even at high
energies is the fact that with LOPES-10
only one polarization direction is measured. 

In addition, 
correlations of further shower parameters have to be investigated 
in more detail to find reasons for low efficiencies, 
even after applying the optimized beam-forming procedures.
Here the direction of the shower axis plays the main role:
By simulations~\cite{2Hueg05} it is expected that the 
emission mechanism in the atmosphere and therefore also the 
radio signal strength depends on the zenith, on the azimuth, 
and as a consequence on the geomagnetic angle.
The latter is the angle between the shower axis and the direction
of the geomagnetic field.
Indeed, detailed analyses of LOPES data have
shown~\cite{Falck05,Horne05a,Petro05} that there are preferred 
directions for enhanced radio signals, or vice versa, there is no 
radio signal detection for specific shower conditions, 
especially at the detection threshold. Again, the fact that 
only one polarization direction is measured aggravates the
interpretation of the data.
In addition, for the present analysis there is a 
selection bias: 
For large distances mainly showers with small zenith
angles are selected due to the fact that the trigger condition 
requires high particle densities inside the KASCADE array, which 
is located in a corner of the Grande array.
Small zenith angles lead to higher particle densities at KASCADE
for same primary energy, and therefore to a worse signal to noise
ratio in the antennas. Further on, the probability to detect 
an air shower in radio might also be
influenced by variations of the background noise 
due to weather conditions or day-night effects in the human 
dominated environment of LOPES. 
In summary, 
events with primary energies even below $10^{17}\,$eV could be
detected in the radio domain, which is remarkably low 
considering the noisy environment at the experimental site and 
the missing measurements of the second polarization direction.

\subsection{Lateral dependence of the received signal} 

After linear scaling of the pulse amplitude $\epsilon_\nu$ 
with the primary energy estimated by KASCADE-Grande a 
clear correlation with the distance is found 
(Fig.~\ref{expomeanr}). 
\begin{figure}[ht]
\begin{center}
\includegraphics[width=6.5cm]{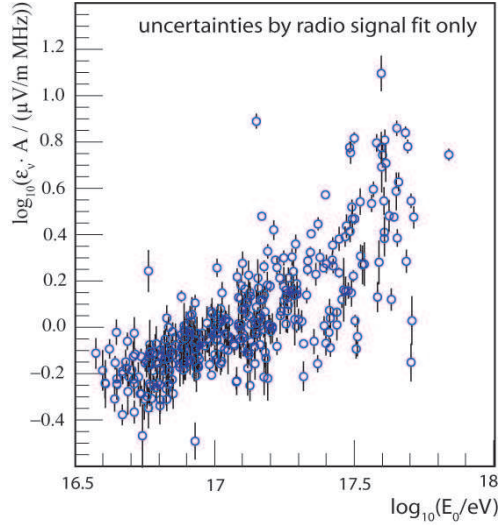}
\end{center}
\caption{Correlation of the pulse height (corrected for the lateral
mean distance of the shower axis to the radio antenna 
system) with the primary energy of the showers.
No correction for the dependence on the geomagnetic angle, 
on the zenith angle, or the azimuth angle are applied.}
\label{energy}
\vspace*{0.3cm}
\end{figure}
The functional form of this dependence and also the lateral 
scaling parameter is of high interest for the further 
development of the radio detection technique.
Following Allan's formula an exponential behavior 
with a scaling parameter of $R_0=110\,$m is expected 
(eqn.\ref{allan} and~\cite{Allan71}) for vertical showers. 
Such an exponential dependence of signal to distance is also 
expected by detailed simulations of the geosynchrotron effect 
with a scaling radius of $\sim 100$ to $\sim 800\,$m, increasing with 
increasing zenith angle~\cite{2Hueg05}. 
The CODALEMA experiment does also support
such a dependence by a preliminary analysis of a few 
events ($R_0$ of a few hundred meters)~\cite{Ardou05a}.
Fitting the present data set (Fig.~\ref{expomeanr})
explicitly assuming an exponential function,  
$R_0$ results to $230\pm51\,$m, i.e. somewhat larger than 
Allan's suggestion which is also drawn in Fig.~\ref{expomeanr}.
One has to note that the initially introduced energy dependent 
selection cut (eqn.~\ref{re0cut}), and the large noise contribution 
(weak signal) for large distances certainly bias the obtained result 
on the lateral scaling parameter towards a flatter slope.
In addition, the selection cut, 
the missing correction to the zenith
angle dependence,
as well as the different 
definition of $R_{\rm mean}$ compared to the definition of the 
distance $R$ used in Allan's formula 
surely distort the obtained scaling parameter.

\subsection{Energy dependence of the received signal} 

In Fig.~\ref{energy} the pulse amplitudes are now scaled 
according to the exponential radial factor obtained by the fit 
described in the previous section, without the 
prior energy correction. 
By that, a clear correlation between the radio field strength and 
the primary energy is found. 

By scaling in addition subsequently with the angle to 
the geomagnetic field, the azimuth, and the zenith angles, 
an even stronger correlation with less fluctuations would be
expected. 
Due to the low statistics and the discussed systematic 
uncertainties of the experimental configuration as well as 
the efficiency behaviour of the detection, these steps 
are omitted in the current phase of the analysis. 
In particular, the small detection efficiencies at 
low energies distort the correlation to a flatter behaviour.
Consequently, we omit also to perform a fit to the 
presented data. Nevertheless, the shown correlation supports 
the expectation that the field strength $\epsilon_\nu$ 
increases by a power-law with an index close to one
with the primary energy, i.e. that 
the received energy of the radio signal increases quadratically 
with the primary energy of the cosmic rays.
The index of this power-law exactly one would 
serve as a proof of the coherence of the radio emission 
during the shower development. 
  
\section{Summary and Outlook}

A combined data analysis  
correlating the radio signals measured by LOPES-10 with extensive 
air shower events reconstructed by KASCADE-Grande was performed. 
The analyzed showers have their axis in up to $700\,$m distance 
from the antenna array and in the primary energy range of 
$10^{16.5}-10^{18}\,$eV. 
Some general dependences of the measured radio signal on certain 
shower parameters were discussed. 
Missing statistics and experimental deficiencies
hamper more detailed investigations in this early stage of air shower 
radio detection experiments.  
Nevertheless, the presented first analysis led to some interesting 
results which can be summarized as follows:
\begin{itemize}
\item 
The most crucial element of radio detection is finding the 
coherence of the radio pulses.
The coherence itself is very sensitive to the shower 
direction and shower core position. On the one hand, 
very small fluctuations in the shower 
observables reconstructed by Grande translate into large fluctuations 
in the estimated radio pulse amplitude. 
On the other hand, by maximizing the radio 
coherence, improved estimations of the core and 
direction parameters can be performed.
Therefore, the maximization of the radio coherence 
(optimized beam-forming) plays a key role in detecting EAS 
radio signals. 
It increases the efficiency of radio detection and improves 
the quality of the correlations between the 
radio signal intensity and the other EAS parameters. 
\item 
LOPES-10 is able to detect radio signals (at $40-80\,$MHz)
induced by extensive air showers even at distances of more 
than $500\,$m from the shower 
axis for primary energies above $10^{17}\,$eV. 
\item
For LOPES-10 an energy detection threshold for primary energies 
below $10^{17}\,$eV was found, which is remarkably low 
considering the noisy environment at the experimental site and 
the missing second polarization measurement. 
\item
The dependence of the radio signal strength on the distance of 
the antennas to the shower axis can be described by an 
exponential function with a scaling radius in the order of a 
few hundred meters.
\item
After scaling with the lateral dependence, a nearly linear increase
of the received field strength with the primary energy could be 
obtained, confirming the coherent character of the emission 
mechanism during the shower development, as expected by 
simulations of the geosynchrotron mechanism.
\end{itemize}
With the measurements of LOPES-10 it could be shown that the 
radio signal in air showers depends on various parameters: 
The primary energy, the distance to the shower axis, and 
the direction of the shower axis. 
Additionally, for all dependences the polarization 
direction of the measurements plays a role, which was not 
investigated with LOPES-10.  
In the present analysis first hints to the lateral structure and 
to the energy dependence of the radio signal could be derived.
With data of the current extension, LOPES-30, consisting of 
thirty absolute calibrated antennas and the possibility of 
an additional trigger by Grande, a larger baseline and higher 
statistics in the 
measurements will be available. 
With LOPES-30 one can also use independent subsets of 
antennas for the 
CC-beam estimate, and would than have the possibility to 
estimate the electric field several times, 
which allows a reconstruction of the 
lateral extension of the radio emission per single air shower.
Additionally, polarization measurements will be performed with
LOPES-30. 
With that data set LOPES-30 is expected to calibrate the 
radio emission in air showers in the primary energy range from 
$10^{16}\,$eV to $10^{18}\,$eV.

{\ack \small
The authors would like to thank the engineering and technical staffs 
of the involved institutes. They contribute with
enthusiasm and commitment to the success of the experiment. 
The corresponding author (A.F.B.) acknowledges very useful 
discussions with Dr.~Tom~Thouw.
LOPES was supported by the German Federal Ministry of Education 
and Research (Verbundforschung Astroteilchenphysik). 
This work is part of the research programme of the
Stichting voor Fundamenteel Onderzoek der Materie (FOM), 
which is financially supported by the Nederlandse Organisatie 
voor Wetenschappelijk Onderzoek (NWO). 
The KASCADE-Grande experiment is
supported by the German Federal Ministry of
Education and Research, the MIUR of Italy,  
the Polish State Committee for Scientific Research 
(KBN grant 1 P03B03926 for 2004-06) and the 
Romanian Ministry of Education and Research (grant CEEX
05-D11-79/2005).
}

\end{document}